\documentclass[12pt,preprint]{aastex}
\slugcomment{}

\shorttitle{Na and O abundances in NGC 2808}
\shortauthors{Carretta et al.}

\begin{document}


\title{Star-to-star Na and O abundance variations along the red giant branch 
in NGC 2808  \altaffilmark{1}}


\author{Eugenio Carretta\altaffilmark{2}, Angela Bragaglia \altaffilmark{2},
Carla Cacciari\altaffilmark{2} }  

\altaffiltext{1}{Based on data collected at the European Southern Observatory,
Chile, during the FLAMES Science Verification.}
\altaffiltext{2}{INAF, Osservatorio Astronomico di Bologna, via Ranzani 1,
        40127,  Bologna,  Italy. carretta@pd.astro.it,
        angela.bragaglia@bo.astro.it, carla.cacciari@bo.astro.it}


\begin{abstract}

We report for the first time Na and O abundances  from high-resolution, high
$S/N$ echelle spectra of 20 red giants in NGC 2808, taken as part of the
Science Verification program of the FLAMES multi-object spectrograph at the ESO
VLT. In these stars, spanning about 3 mag from the red giant branch (RGB) tip,
large variations are detected in the abundances of oxygen and sodium,
anticorrelated with each other; this is a well known evidence of proton-capture
reactions at high temperatures in the ON and NeNa cycles.  One star appears
super O-poor; if the extension of the Na-O anticorrelation is confirmed, NGC
2808 might reach O depletion levels as large as those of M 13. This result
confirms our previous findings based on lower resolution spectra (Carretta et
al. 2003) of a large star-to-star scatter in proton capture elements at all
positions along the  RGB in NGC 2808, with no significant evolutionary
contribution. Finally, the average metallicity for NGC 2808 is [Fe/H]$=-1.14
\pm 0.01$ dex ($rms=0.06$) from 19 stars.

\end{abstract}


\section{Introduction}

Since the pioneering work of Osborn (1971), a large body of evidence  has
accumulated indicating that globular clusters (GCs) are $not$ mono-metallic
populations, as far as the light elements  (C, N, O, Na, Mg, Al) are concerned.

While early studies on the light elements C, N, and carbon isotopic ratios 
showed that some kind of mixing was involved in RGB stars (C abundances 
declining with increasing stellar luminosities, C and N abundances
anticorrelation among stars of same evolutionary phases, see Smith 1987 for a
review), they could not explain why these variations were observed all the way
down to unevolved stars where mixing is not supposed to play any role (e.g. the
recent studies by Cannon et al. 1998; Harbeck et al. 2003; and references
therein). 

The Lick-Texas study of heavier elements (see Kraft 1994 and Sneden et al.
2004: S04, for complete references) in bright giants in several clusters showed
that the [Na/Fe] and [O/Fe] ratios were anticorrelated:  O-depletion was always
accompanied by Na-enhancement,  the stars in M 13 displaying the most severe
variations. Whenever C and N abundances were available, N was anticorrelated
with O, whereas a clear correlation existed between O and C abundances. This
points to a redistribution of C,N,O in a H-burning CNO-cycle. The theoretical
background presented by Denisenkov \& Denisenkova (1990)  and Langer et al.
(1993) clarified that proton-capture reactions at high temperatures in the ON
and NeNa cycles are involved in building up the observed pattern.

While these anomalies are markedly confined to the dense cluster environment
(e.g. Gratton et al. 2000), the true site where  CNO and NeNa cycles concur to
form the observed abundance pattern is still a  matter of debate. In fact, the
same chains of p-captures (on C,N,O,Ne,Mg) may occur both in the H-burning
shell of low mass ($< 1$ M$_\odot$) stars presently climbing up the RGB branch,
and in the so called Hot Bottom Burning (HBB: Bl\"ocker \& Sch\"onberner 1991,
Boothroyd \& Sackmann 1992) taking place in a prior generation of
intermediate mass ($3-8$ M$_\odot$) stars in the asymptotic giant branch
(IM-AGB) phase. 

The first direct observations of O and Na abundances in dwarfs at the
main-sequence turn-off in NGC 6752 (Gratton et al. 2001) and in 47 Tuc
(Carretta et al. 2004) are a clearcut evidence that part of the observed Na-O
anticorrelation must be primordially established, since these unevolved stars
do not have either high enough temperatures in their cores for the required
reactions or large enough convective envelopes to dredge up to the surface the
products of p-capture reactions.

Here we present the results for Na and O obtained for 20 RGB stars in NGC 2808,
observed during the FLAMES Science Verification.  Carretta et al. (2003)
exploited the MEDUSA mode of FLAMES at VLT-UT2 to uncover large star-to-star
variations in Na abundances among 80 RGB stars in this cluster. In this Letter,
we show that Na is anti-correlated with O, and we find  preliminary evidence
that in NGC 2808 the O-depletion in some stars could be as extreme as in M 13
super-O-poor stars.

\section{Atmospheric parameters and abundance analysis}

\begin{figure}
\plotone{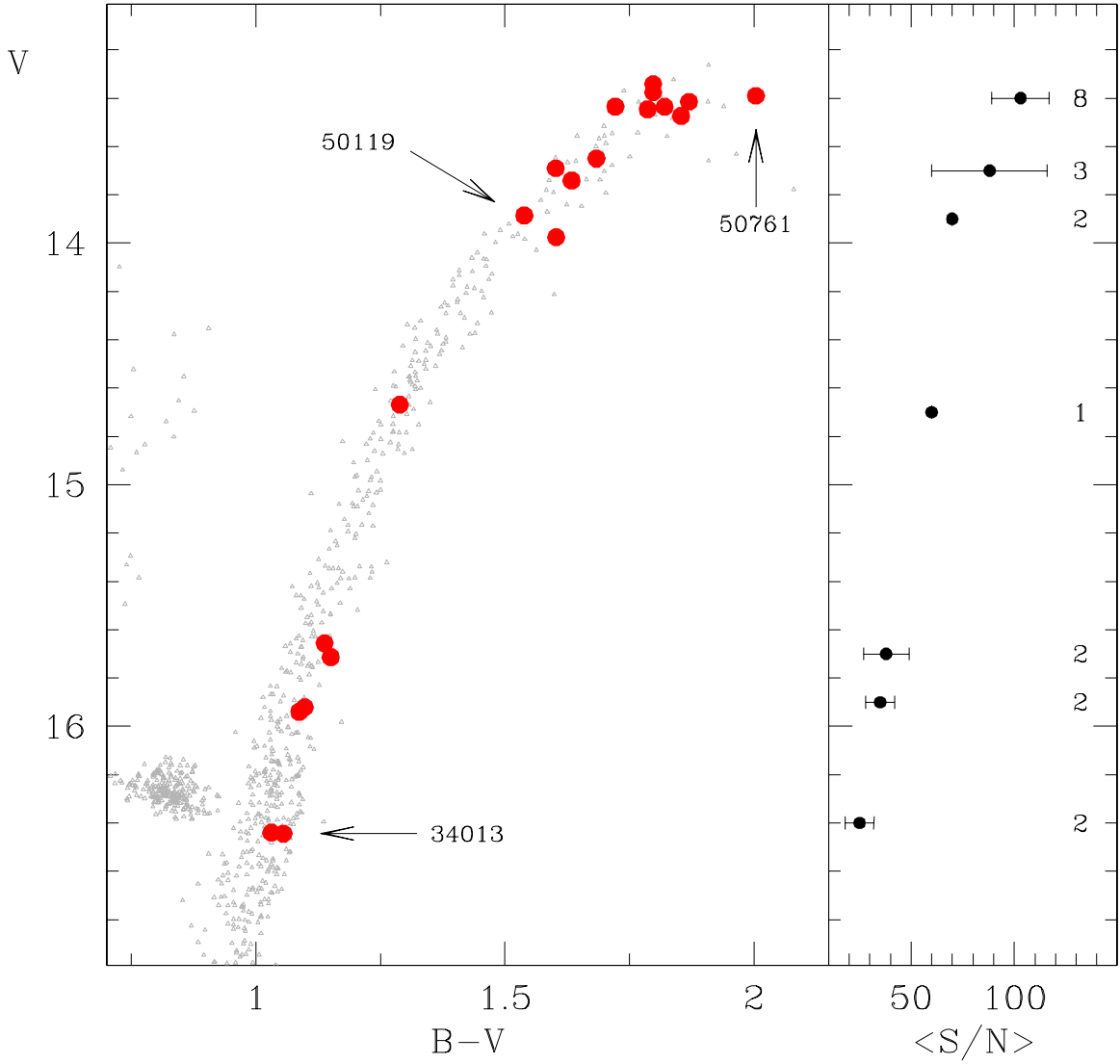}
\caption{Observed stars are indicated by larger filled circles on the $V,B-V$
CMD of NGC 2808 (Piotto et al., priv.comm.). The right panel shows the average
$S/N$ at different magnitude levels, and numbers indicate how many stars were
averaged each time. The coolest star is indicated, as well as the ones with the
lowest $S/N$ (34013, dropped from discussion) and with the lowest O abundance
(50119).
\label{cmd}}
\end{figure}

Twenty RGB stars with $V=13.2-16.5$  were observed with the fiber-fed UVES Red
Arm  ($R=47000$, spectral coverage of 200 nm, centered at 580 nm). Full details
of the observations and data reduction, magnitudes, and coordinates  can be
found in Cacciari et al. (2004; C04). The $S/N$ ratio varies a lot (the
selection was optimized for studying mass loss, not for abundance  analysis)
and is shown in Figure~\ref{cmd} together with the target positions in the CMD.

Effective temperatures T$_{\rm eff}$ (from colors and the Alonso et al. 1999
calibration) are discussed in C04 and
Carretta et al. (2003). Surface gravities $\log g$ were obtained from
temperatures and bolometric corrections, using the distance modulus
$(m-M)_V=15.59$ (Harris 1996), and assuming that the stars have masses of 0.85
M$_\odot$. The adopted  bolometric magnitude of the Sun is $M(bol)_\odot =
4.75$.  Equivalent widths ($EW$s) for Fe I, Fe II and Na lines were measured as
described in Bragaglia et al. (2001). Details will be presented in a
forthcoming paper.  We derived the microturbulence velocity $v_t$ by zeroing
the slope of the abundances of Fe I vs $EW$s and the overall model metallicity
was chosen using the Kurucz (1995) grid of model atmospheres with the
overshooting option set on. Abundances of Fe I, Fe II and Na I were derived
from the analysis of $EW$s (for Na from the doublets at 5682-88~\AA\ and
6154-60~\AA). O abundances were derived from spectral synthesis of the
forbidden lines [O I] at 6300.31 and 6363.79~\AA, after cleaning the observed
spectra for telluric line contamination. The contribution of the weak, high
excitation Ni line at 6300.34~\AA, with the laboratory $\log gf$ recently
measured by Johansson et al. (2003), is neglibile (about 0.5 to 1.2 m~\AA) in 
the entire magnitude range. Reference solar abundances and atomic parameters
for lines are those of Gratton et al. (2003). Adopted parameters and derived
abundances are listed in Table 1.

The ionization equilibrium FeII/FeI is good: on average, [Fe/H]II-[Fe/H]I$ =
0.00 \pm 0.03, rms=0.15$ dex (19 stars: for star 34013, with $S/N\sim 20$, 
measurements of $EW$s were unreliable and this star was dropped from further
discussion). Excluding the coolest star (50761), there seems to be a slight
trend for increasing the FeII-FeI difference with decreasing T$_{\rm eff}$.
While this could hint to possible departures from LTE (overionization), the
good agreement of Fe I abundances all over the sampled range in T$_{\rm eff}$
argues against an underestimate of [Fe/H] abundances. Another possibility is
that the structure of the atmospheres for coolest giants is not well
reproduced by models of the Kurucz grid (see Dalle Ore 1993).
No noticeable trend of derived abundances as a function of the excitation
potential is discernible. This supports the adoption of the Alonso et al.
temperature scale, as found also by Ivans et al. (2001). Estimates of
typical errors in T$_{\rm eff}$, $\log g$, [A/H] and $v_t$ are 70 K, 0.1 dex,
0.1 dex and 0.1 km s$^{-1}$, respectively. When combined to errors in the
measurement of $EW$s, they translate into a total (internal) error of about 0.08
dex in [Fe/H] and [Na/Fe] and 0.13 dex in [O/Fe].

Average values for NGC 2808 are [Fe/H]I$=-1.14 \pm 0.01, rms=0.06$ dex and 
[Fe/H]II$=-1.14 \pm 0.03, rms=0.13$ dex. This is the first determination of the
iron abundance based on modern high resolution spectra for this cluster.

\section{Results and discussion}

\begin{figure}
\epsscale{0.6}
\plotone{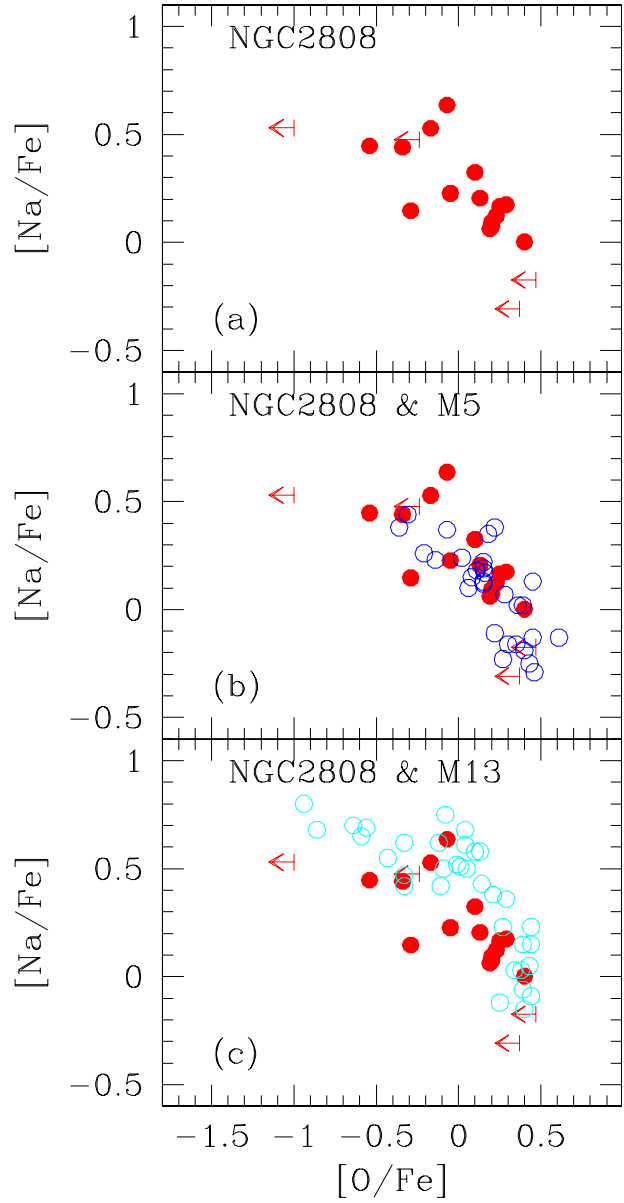}
\caption{[Na/Fe] vs [O/Fe] ratios in red giants of NGC 2808 (upper panel).
Filled circles are effective detections. Upper limits in O are indicated by
arrows. Middle panel: the same, with overimposed RGB stars in M 5 (Ivans et
al. 2001; open circles). Lower panel: the same, with overimposed RGB stars 
in M 13 (Sneden et al. 2004; open circles).
\label{anti}}
\end{figure}

Derived abundances of Na and O are plotted in Figure~\ref{anti}.
[O/Fe] ratios were referred to Fe II; [Na/Fe]
ratios are corrected for departures from LTE as in Gratton et al.
(1999), and are referred to Fe I abundances. This is the first
study to unveil the existence of the Na-O anticorrelation in this cluster.
In Carretta et al. (2003), we pointed out the large spread observed in Na
abundances at all luminosities along the RGB; 
however, the MEDUSA spectra (acquired to study the mass loss) were centered on
the Na D or H$\alpha$  features and they did not contain O lines.

\begin{figure}
\epsscale{0.6}
\plotone{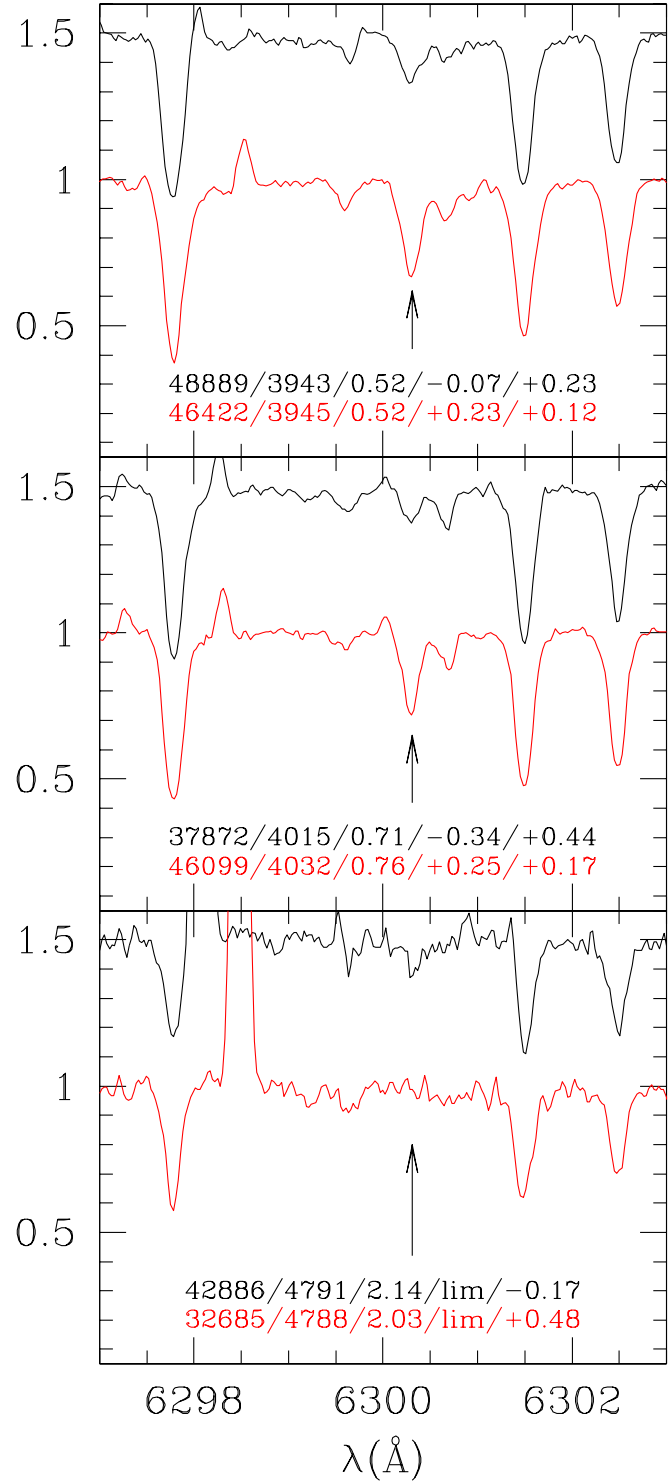}
\caption{Comparison of observed spectra of 3 pairs of RGB stars in NGC 2808.
Arrows indicate the [O I] 6300.31~\AA\ forbidden line. Each pair has 
similar atmospheric parameters, as indicated by the labels, that give star id,
T$_{\rm eff}$, $\log g$, [O/Fe] and [Na/Fe].
\label{spettri}}
\end{figure}

Here we show that also O abundances are very different among
stars at the same position along the RGB. We show in
Figure~\ref{spettri} the strengths of the forbidden [O I] line at 6300~\AA\ in
3 pairs of stars with similar atmospheric parameters, yet quite large
difference in the Na and O content, from just above the RGB-bump
(where only upper limits could be derived for O; nevertheless, stars 42886 and
32685 differ by  0.5 dex in Na abundances), to the upper RGB and RGB tip.
From this Figure it is clear that large star-to-star scatter in $both$ O and
Na abundances does exist among red giants in NGC 2808.

In Figure~\ref{anti} we also compare our data in NGC 2808 to similar ones for M
5 (Ivans et al. 2001) and for M 13 (S04), after an adjustment for the offset in
the adopted  solar O abundance (0.14 dex). The comparison with M 5 (a cluster
with almost the same mean [Fe/H] as NGC 2808, see e.g. Carretta and Gratton
1997) shows that the Na-O distributions are grossly similar. In the high-O,
low-Na regime, stars observed by Ivans et al. (2001) have mostly spectra with
$S/N>60$, whereas only a $S/N\sim 30$ could be reached for our 2 stars of NGC
2808, giving only upper limits for O abundances. However, the impression is
that in the O-poor, Na-rich part of this diagram the two distributions might be
somewhat different, with NGC 2808 reaching larger O depletions.
For the most O-poor star (50119), even with $S/N \simeq 70$,  the [O I] line at
6300.31~\AA\ is vanishingly small and only an upper limit could be assigned. 
The comparison with M 13, the archetype for extremely O-depleted stars, seems
to show that a similar degree of O-depletion is possibly reached also in NGC
2808. Admittedly, this is based on only one star. Higher quality spectra,
purposedly acquired, should be taken to verify this issue.

The recent finding of a clear Na-O anticorrelation among unevolved stars in
NGC 6752 (Gratton et al. 2001) and in 47 Tuc (Carretta et al. 2004) strongly
points out that variations in these elements must be due to H-burning at
high temperature in stars able to dredge-up and then eject their O-poor, 
Na-rich matter. In fact, because dwarf stars are unable to manage either 
of these requirements, an external origin for the Na and O abundance pattern 
must be necessarily accepted. IM-AGB stars are found to be good 
candidates (see e.g. Ventura et al. 2002).

A certain amount of further modifications as the star climbs up the RGB was
postulated, since the lightest elements, like carbon, show a decrease as a 
function of stellar luminosity. For Na and O, the only clear 
evidence was found for M 13, where a shift in the average abundances was noted
for the brightest red giants (Pilachowski et al. 1996; Kraft et al. 1997), 
above $\log g \sim 1.0$.
However, recently S04 found that the anticorrelation between O and Mg  isotopic
ratios in M 13 is very similar to that in NGC 6752 (Yong et al. 2003), where
the primordial pattern of chemical anomalies is well established. On this
basis, S04 concluded that also in M 13 the extreme O-depletions are likely
originated in IM-AGB stars.  Apparently, the observed Na-O pattern in NGC 2808
is not very different from the marked anti-correlation in M 13. The
implication is that also in NGC 2808 we are seeing the results of a pattern of
abundances already established by a first generation of IM-AGB stars
(see e.g., Parmentier 2004; D'Antona \& Caloi 2004).
An evolutionary contribution, if any,
is nevertheless very small in this cluster (Carretta et al. 2003).

Is there any link between chemical inhomogeneities and global properties in
GCs? This remains presently an unsolved issue, but this study in NGC
2808 may allow to add another piece to the puzzle. In fact, NGC 2808 and M 13 
have a rather different HB morphology, in spite of their similar [Fe/H].
NGC 2808 is
the most famous example of bi-modal HB: almost $1/4$ of its HB stars lies in
the blue part of the CMD, while presenting a well populated, stubby red HB.
Yet, our findings seem to show O-depletions as large as in M 13, where only a
long blue HB is seen. On the other hand, NGC 6752 (whose blue HB is very
similar to the one of M 13) does not show (Yong et al. 2003) super O-poor stars
as observed in M 13 and, likely, in NGC 2808. 

Clusters like M 5 and M 3 are, in this respect, more similar to NGC 6752. The
Na-O anticorrelation in these clusters does not reach 
the extreme values of M 13 (S04; Ivans et al. 2001). Moreover, NGC 6752 is a
post-core collapse system (Harris 1996), whereas the other quoted GCs have
different, but not $extremely$ different, concentration parameters. Still,
variations in Na and O are observed also in the much less massive and loose
cluster Pal 5 (Smith et al. 2002), presently leaving a trail of its stars into
the galactic field due to disruptive processes like tidal shocking.

In summary, abundance inhomogeneities in elements  produced by p-capture
fusions at high temperature are observed in GCs independently of 
(i) total mass and
concentration, 
(ii) galactic (disk or halo) population, 
(iii)  overall metallicity, 
and (iv) HB morphological type. 
At face value, the Na-O anticorrelation is found in every
cluster surveyed insofar, despite the large differences in global properties. 
Following Occam's razor, this might imply that we are seeing the outcome of a
process intimately related to dense aggregates, maybe intrinsically connected
to their own formation process. This is supported by the lack
of such a behaviour in field stars (see Gratton et al. 2000). 

What to do next? We are quite confident about the $shape$ and meaning of the
Na-O anticorrelation in term of products of p-capture reactions in the ON and
NeNa cycles. A step forward would
be now to study the {\it distribution function} of the anticorrelation, 
using high quality spectra of large samples ($\sim$ 100) of
stars along the RGB in several clusters spanning a variety of parameters (age,
metallicity, HB morphology, etc.), to look for any possible connection
unnoticed until now due to the paucity of samples. This is a project we are
currently working on using the FLAMES multi-object spectrograph at ESO-VLT.

\acknowledgments
The authors thank R.G. Gratton for invaluable suggestions and discussion, 
the ESO staff for the
observations and the preliminary data reduction, and G. Mulas for performing a
more detailed data reduction.

\clearpage

\begin{deluxetable}{rrcccrccrcclllr}
\tabletypesize{\scriptsize}
\tablecaption{Adopted parameters and derived abundances.\label{parabu}}
\tablehead{\colhead{Star ID}&
\colhead{$S/N$}&
\colhead{T$_{\rm eff}$}&
\colhead{$\log g$}&
\colhead{$v_t$}&
\colhead{n}&
\colhead{[Fe/H]I}&
\colhead{$\sigma$}&
\colhead{n}&
\colhead{[Fe/H]II}&
\colhead{$\sigma$}&
\colhead{n}&
\colhead{[Na/Fe]}&
\colhead{$\sigma$}&
\colhead{[O/Fe]}\\
\colhead{}&
\colhead{}&
\colhead{K}&
\colhead{dex}&
\colhead{km s$^{-1}$}&
\colhead{}&
\colhead{dex}&
\colhead{dex}&
\colhead{}&
\colhead{dex}&
\colhead{dex}&
\colhead{}&
\colhead{dex}&
\colhead{dex}&
\colhead{dex}\\
}
\startdata
%#star     S/N  Teff logg  Vt      nr  [Fe/H]I rms    nr [Fe/H]II rms  n   [Na/Fe] rms    [O/Fe]
%#              V-K  V-K 	         fin 	             KS
10201 &   45 & 4717 &2.02 &1.20 & 51 &$-$1.06 &0.13&  8&$-$1.21& 0.05& 3&  +0.23& 0.06 & $-$0.05 \\ %u
13983 &   40 & 4826 &2.17 &0.60 & 33 &$-$1.08 &0.08&  5&$-$1.09& 0.13& 2&  +0.45& 0.12 & $-$0.54 \\ %s
32685 &   30 & 4788 &2.03 &0.83 & 34 &$-$1.15 &0.09&  5&$-$1.41& 0.07& 2&  +0.48& 0.10 &$<-$0.24 \\ %x
34013 &   20 & 5110 &2.51 &0.80 & 29 &$-$0.89:&0.21&   &       &     &  &       &      &         \\ %v
37872 &  120 & 4015 &0.71 &1.68 & 96 &$-$1.10 &0.11& 14&$-$1.06& 0.08& 4&  +0.44& 0.07 & $-$0.34 \\ %i
42886 &   30 & 4791 &2.14 &0.85 & 47 &$-$1.16 &0.14&  4&$-$1.38& 0.10& 2&$-$0.17& 0.13 &$<+$0.47 \\ %w
43217 &   30 & 4916 &2.41 &0.80 & 33 &$-$1.00 &0.12&  7&$-$1.17& 0.13& 1&$-$0.31&      &$<+$0.37 \\ %r
46099 &   80 & 4032 &0.76 &1.72 & 60 &$-$1.18 &0.12& 10&$-$1.19& 0.12& 2&  +0.17& 0.03 &   +0.25 \\ %o
46422 &  100 & 3943 &0.52 &1.85 & 95 &$-$1.17 &0.12& 14&$-$1.08& 0.09& 3&  +0.12& 0.11 &   +0.23 \\ %n
46580 &   65 & 4051 &0.74 &1.68 & 89 &$-$1.15 &0.11& 15&$-$1.03& 0.13& 4&  +0.33& 0.05 &   +0.10 \\ %e
47606 &  110 & 3839 &0.44 &1.66 & 85 &$-$1.12 &0.14& 14&$-$1.14& 0.16& 4&  +0.09& 0.05 &   +0.20 \\ %c
48609 &  110 & 3846 &0.44 &1.78 & 58 &$-$1.22 &0.12&  8&$-$1.11& 0.13& 3&  +0.07& 0.09 &   +0.20 \\ %f
48889 &   85 & 3943 &0.52 &1.80 & 99 &$-$1.15 &0.15& 13&$-$1.22& 0.12& 4&  +0.64& 0.13 & $-$0.07 \\ %a
50119 &   70 & 4166 &0.93 &1.73 &119 &$-$1.08 &0.15& 18&$-$1.21& 0.13& 4&  +0.53& 0.03 &$<-$1.00 \\ %m
50761 &  120 & 3756 &0.31 &1.75 & 81 &$-$1.22 &0.12& 10&$-$0.81& 0.07& 4&  +0.15& 0.12 & $-$0.29 \\ %h
51454 &  120 & 3893 &0.51 &1.65 & 90 &$-$1.26 &0.11& 14&$-$1.11& 0.13& 4&  +0.21& 0.07 &   +0.13 \\ %l
51499 &   85 & 3960 &0.57 &1.70 &104 &$-$1.25 &0.10& 18&$-$1.24& 0.12& 4&  +0.17& 0.10 &   +0.29 \\ %d
51983 &   95 & 3855 &0.47 &1.77 & 71 &$-$1.15 &0.11& 13&$-$1.16& 0.12& 4&  +0.53& 0.09 & $-$0.17 \\ %b
53390 &   60 & 4426 &1.43 &1.30 &106 &$-$1.12 &0.10& 19&$-$1.14& 0.13& 4&  +0.00& 0.09 &   +0.40 \\ %q
56032 &   70 & 4045 &0.87 &1.70 &103 &$-$1.10 &0.10& 17&$-$0.99& 0.12& 4&  +0.06& 0.09 &   +0.19 \\ %y
\enddata
\tablenotetext{a}{n is the number of measured lines.}
\end{deluxetable} 

\clearpage

\clearpage

\clearpage

\end{document}